\documentclass[%
superscriptaddress,
reprint,
 amsmath,amssymb, amsthm, aps, prl,
]{revtex4-1}
\usepackage[colorlinks,bookmarks=false]{hyperref} 
\usepackage{graphicx}
\usepackage{dcolumn}
\usepackage{bm}
\usepackage[titletoc,toc,title]{appendix}
\usepackage{enumitem}
\usepackage{dsfont}
\usepackage{color}
\usepackage[normalem]{ulem} 
\usepackage{times}
\usepackage{latexsym}
\usepackage{amsfonts}
\usepackage{pifont}
\usepackage{stmaryrd}
\usepackage{array}
\usepackage{epsfig}
\usepackage{verbatim}
\usepackage{multirow}
\usepackage{rotating}
\usepackage{enumerate}
\usepackage{color}
\usepackage[dvipsnames]{xcolor}
\usepackage{comment}

\usepackage{sidecap}

\def\f12{\frac{1}{2}}

\def\kin{\ket{\mathrm{in}}}
\def\kout{\ket{\mathrm{out}}}

\newcommand {\ket}[1]{\lvert \, #1\rangle}
\newcommand {\bra}[1]{\langle #1 \, \rvert}
\newcommand {\braket}[2]{\langle #1 \, | \, #2 \rangle}

\newcommand{\be}{\begin{equation}}
\newcommand{\ee}[1]{\label{#1} \end{equation}}
\newcommand{\bbe}{\begin{equation*}}
\newcommand{\eee}{\end{equation*}}

\newcommand{\bs}{\begin{split}}
\newcommand{\es}{\end{split}}

\def\f12{\frac{1}{2}}

\begin{document}
\author{Magdalena Zych}\email{m.zych@uq.edu.au}
\affiliation{Centre for Engineered Quantum Systems, School of Mathematics and Physics, The University of Queensland, St Lucia, QLD 4072, Australia}
\author{Fabio Costa}
\affiliation{Centre for Engineered Quantum Systems, School of Mathematics and Physics, The University of Queensland, St Lucia, QLD 4072, Australia}
\author{Timothy C. Ralph}
\affiliation{Centre for Quantum Computation and Communication Technology,
School of Mathematics and Physics, The University of Queensland, St Lucia, QLD 4072, Australia }

\title{Relativity of quantum superpositions}
\begin{abstract}
In modern physics only relative quantities are considered to have physical significance.  For example, position assigned to a system depends on the choice of coordinates, and only relative distances between different systems have physical relevance.
On the other hand, in quantum theory the scenario where one system, A,  is localised around certain position while another system B is in a spatial superposition is considered to be physically different from the scenario where A is in a superposition, while B is localised. Different physical effects are anticipated in the two scenarios especially when the two systems have widely different masses.  Here we show that for any superposed amplitudes that are related by a symmetry transformation, the above scenarios are physically equivalent: probabilities for any measurement are the same whether we assign the superposition state to system A or to system B. More generally, linearity of quantum theory guarantees that if a theory is invariant under some symmetry transformations it is necessarily also invariant under their arbitrary ``superpositions''. Thus the notion of a superposition turns out to be relative to the choice of coordinates,  once it is realised that relations between coordinates do not need to be classical.  
\end{abstract}
\maketitle
\paragraph{Introduction.--}
General covariance demands that absolute position has no physical meaning \cite{WeinbergGR}. A scenario where one particle is localised around some position $x_1$ and a second particle is at $x_2=x_1+L$, in any translationally invariant theory, is exactly the same as the scenario where the first particle is localised around some different point $x'_1$,  while the other is at  $x'_2=x'_1+L$. The difference between the two is considered to be only apparent, and is attributed to a different choice of a coordinate system in which the positions of the particles are described. 
More broadly, in modern physics only relative quantities are considered to have physical significance \cite{Gabbay2006philosophy} as a consequence of fundamental interactions being local and depending only on relationships between the systems. 

On the other hand, the following two scenarios are considered to be fundamentally different: In scenario $A$  particle $1$ is localised around  some position $x_1$ while particle $2$ is in a superposition of two states, each localised around  a different position $\ket{x_1+L}+\ket{x_1+L'}$.  
In scenario B  particle $1$ is in  superposition $\ket{x_2-L}+\ket{x_2-L'}$ while particle $2$ is localised around $x_2$. 
If the mass of particle $1$ is much larger than the mass of particle $2$, some authors question whether scenario $B$ can be consistently described with current physical theories \cite{ref:Diosi1989, ref:Penrose1996, Anastopoulos:2015zta, Stamp2015, kiefer2012quantum}. 
Some contest that scenario $B$ is physical at all, postulating that the superposition principle in case  $B$  breaks down at a very short time-scale, while it effectively does not break down in case $A$, see e.g.~\cite{ref:Bassi2013, carney2018massive}. 

The aim of the present work is to investigate what the standard quantum framework tells us about such scenarios. Our main result is a proof that scenarios like $A$ and $B$ above are fully equivalent according to standard quantum theory. Maintaining the view that there is a fundamental difference between them is incompatible with translational invariance. We generalise the result to any symmetry of dynamics and show that linearity of quantum theory guarantees that if dynamics is invariant under certain transformations it is also invariant  under their arbitrary ``superpositions''. 

Importantly, our results do not require the inclusion of the reference frames 
among the dynamical systems, which is at the core of the ``quantum reference frames'' (QRFs) approach \cite{PhysRevD.30.368, Rovelli:1991QRS, RevModPhys.79.555, AngeloPhysics2011, Angelo:2015PRA, Loveridge2018}. The studies of QRFs show that if reference frames are treated as quantum systems, they can be related by non-classical transformations, e.g.~superpositions of Galilean boosts or of translations \cite{Giacomini:2017qcovariance}. Our work is concerned with a very different question: If labelling of the states is a convention (e.g.~stemming from an arbitrary choice of coordinates) does this extend to conventions which yield sets of states that cannot be related by any classical coordinate transformation? Our approach answers this question affirmatively and shows that the resulting quantum transformations between coordinates are still compatible with the usual view of coordinate systems as auxiliary tools, not part of the dynamical system. 
 


\paragraph{Closed box thought experiments.--}\label{sec:closedbox}
Here we formulate a thought experiment demonstrating that it is in principle possible to prepare a state of two systems and certify that their relative distance is in superposition, yet it is not possible to tell which of the two systems is in superposition relative to an external reference. 

Consider a `closed box' containing a massive body with mass $M$ and an atom with a much smaller mass $m$. The box is sufficiently large so that any fields outside are approximately the same for the different configurations of the systems inside. 
A photon with local frequency $\omega$ can be used by an outside agent to interact/measure the systems inside. Before entering the box the photon passes through a beam-splitter, preparing it in a superposition of modes $\kin$, which enters the box,  and $\kout$ which remains outside. 

\begin{figure}[h!]
\includegraphics[width=0.99\columnwidth]{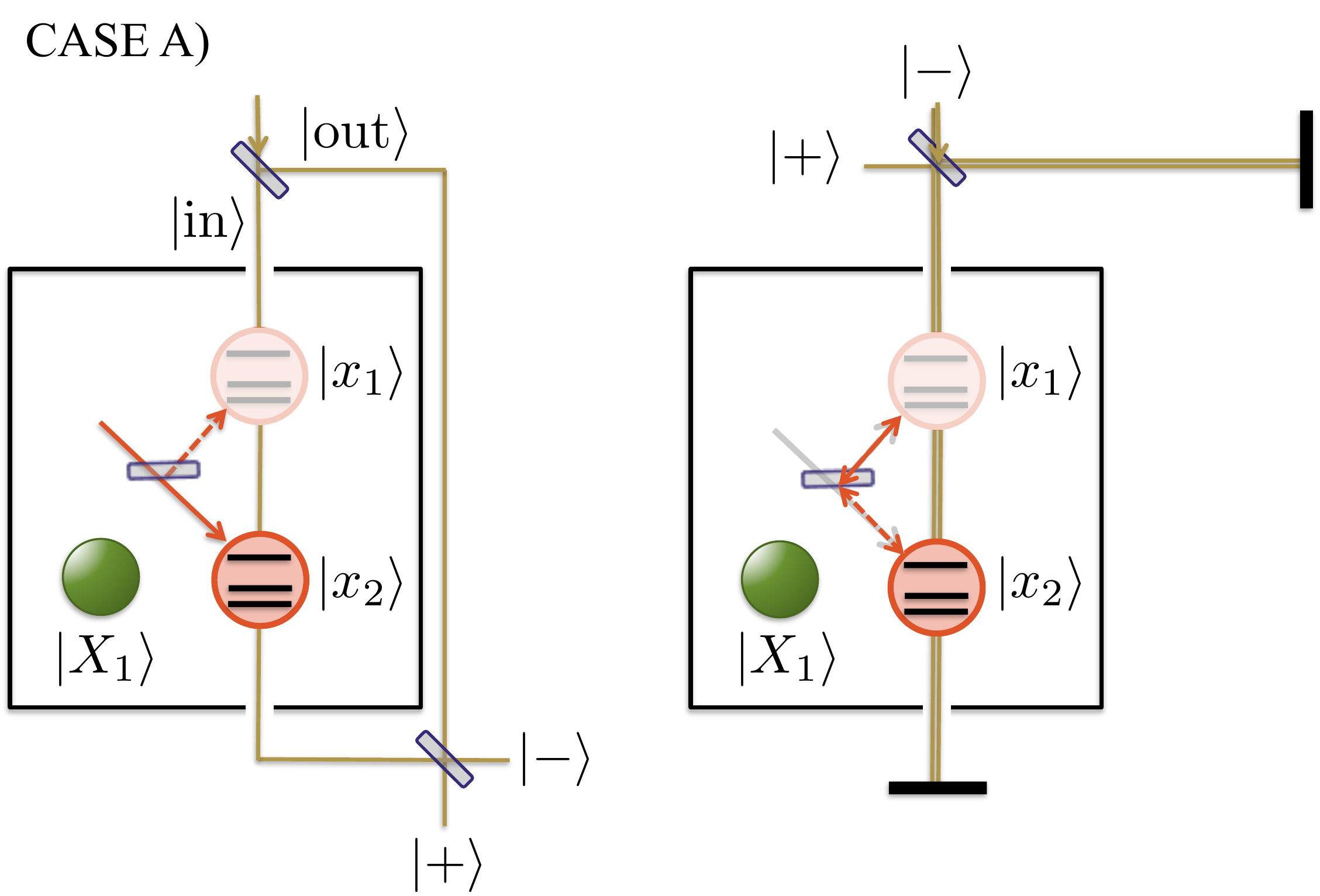}
\caption{Atom in superposition of two distances from a massive body. An outside agent can make measurements on the photon in a superpositions basis $\ket{\pm}\propto\kin\pm\kout$.  The scheme on the left panel allows the outside agent to prepare the atom in one of its two orthogonal states. The scheme on the right panel  allows them to measure the relative phase between the superposed amplitudes of the atom. \hspace*{\fill}}
\label{caseA}
\end{figure}

In case A,  Fig.~\ref{caseA}, the mass $M$ is semi-classically localised and can be assigned state $\ket{X_1}$. The atom passes through a beam-splitter and is thus prepared in superposition of two semi-classically localised states $\ket{x_1}$ and $\ket{x_2}$, at a distance $d_1$ and $d_2$, respectively, from $M$. 
The initial state of the mass, atom and photon reads $\ket{X_1}(\ket{x_1}+ \ket{x_2})(\ket{\mathrm{in}}+\ket{\mathrm{out}})/2$.  
The photon mode in the box is absorbed by the atom. In the local rest frame of the atom at a distance $d_i$ to $M$, to lowest post-Newtonian order, the photon's frequency will be blue-shifted as $\omega(1+GM/d_ic^2)$ \cite{PoundRebka:1960,WeinbergGR}. Thus, we consider an atom with at least three energy levels with the spacing so that it can absorb the photon in both locations $d_i$. After proper time $\tau$ the photon is re-emitted. Due to a gravitational potential difference $\Delta\Phi=GM(1/d_2-1/d_1)$ \cite{cow, TinoFermiSchool, asenbaum2016phase}  between the locations of the atom, a relative phase \cite{Zych:2012} $\Delta\varphi=\omega \Delta \Phi\tau/c^2$ (up to order $1/c^2$) 
is eventually acquired between the photon amplitudes. (The free propagation phase acquired by the photon before and after its interaction with the atom, including the Shapiro effect \cite{ref:Shapiro1964},  is here the same for both amplitudes of the atom.)  For simplicity we assume $\Delta\varphi=\pi$, see Supplemental Material for a general case.   The state of all systems after the photon leaves the box reads 
$\ket{X_1}\big[\left(\ket{x_1}+e^{i\theta}\ket{x_2}\right)e^{i\phi_c}\kout+\left(\ket{x_1}-e^{i\theta}\ket{x_2}\right)\kin\big]/2$.
Here, $\theta= m\Delta\Phi t/\hbar$ is the relative phase between the two amplitudes of the atom due to mass $M$,  $t$ the duration of the experiment as measured outside the box, and $\phi_c$ is the total phase acquired by mode $\kout$,  which can be varied by the outside agent. 

An outside agent can now overlap the photon modes on a beam-splitter,  {creating a Mach-Zehnder interferometer for the photon,} Fig.~\ref{caseA} left panel. The final state (up to a normalisation factor $1/2\sqrt{2}$) reads 
\be
 \begin{split}
\ket{X_1}\big[&\big((1+e^{i\phi_c})\ket{x_1}-e^{i\theta}(1-e^{i\phi_c})\ket{x_2}\big)\ket{+}\\
+&\big((1-e^{i\phi_c})\ket{x_1}-e^{i\theta}(1+e^{i\phi_c})\ket{x_2}\big)\ket{-}\big],
 \end{split}
\ee{caseAfin}
where $\ket{\pm}\propto \kin\pm\kout$ are the two output modes of the interferometer. 
Eq.~\eqref{caseAfin} describes the atom and the photon as entangled. The measurement on the photon in $\ket\pm$ basis yields the two possible outcomes with equal probabilities -- independently of the controllable phase $\phi_c$ -- and prepares the atom in one of its two orthogonal states, as per eq.~\eqref{caseAfin}.
  
The scheme can be extended to yield the measurement of the atom in a superposition basis. This is achieved by performing a Michelson-Morley interferometry with the photon, Fig.~\ref{caseA} right panel. 
{The atom passes again through the beam splitter, which creates a new superposition between $\ket{x_1}$ and $\ket{x_2}$, and the photon modes are reflected back through the box. 
As the photon interacts with the new state of the atom,} the  phases $\theta$ and $\Delta\varphi$ are acquired again by the relevant amplitudes, since they only depend on relative distances between the systems. {The photon modes are overlapped on the beam splitter outside the box and measured in the $\ket\pm$ basis. The final state reads
\be
\frac{1}{2}e^{i\frac{\theta}{2}}\ket{X_1}\sum_\pm\left(J_{1}^{\pm}\ket{x_1}-i e^{i\theta}J_{2}^{\pm}\ket{x_2}\right)\ket{\pm},
\ee{caseAfinMM_pi}
where 
$J_1^\pm=-i\sin(\frac{\theta}{2})\pm e^{i2\phi_c}\!\cos(\frac{\theta}{2})$, and
$J_2^\pm=-i\cos(\frac{\theta}{2})\pm e^{i2\phi_c}\sin(\frac{\theta}{2})$ and yields the photon detection} probabilities $P_\pm=\big(1\mp\sin(\theta)\sin(2\phi_c)\big)/2$, see Supplemental 
Material.
The interference visibility $\left|\sin(\theta)\right|$ depends on the relative phase $\theta$ between the amplitudes of the atom. Since it depends on the total time the atom spends in superposition, the outside agent can vary $\theta$ by changing the length of the photon paths paths outside the box, so as to extend the duration of the experiment.
In summary, the outside agent can confirm that the system inside maintains coherence and can measure the relative phase between the superposed amplitudes.

 

\begin{figure}[h!]
\includegraphics[width=0.99\columnwidth]{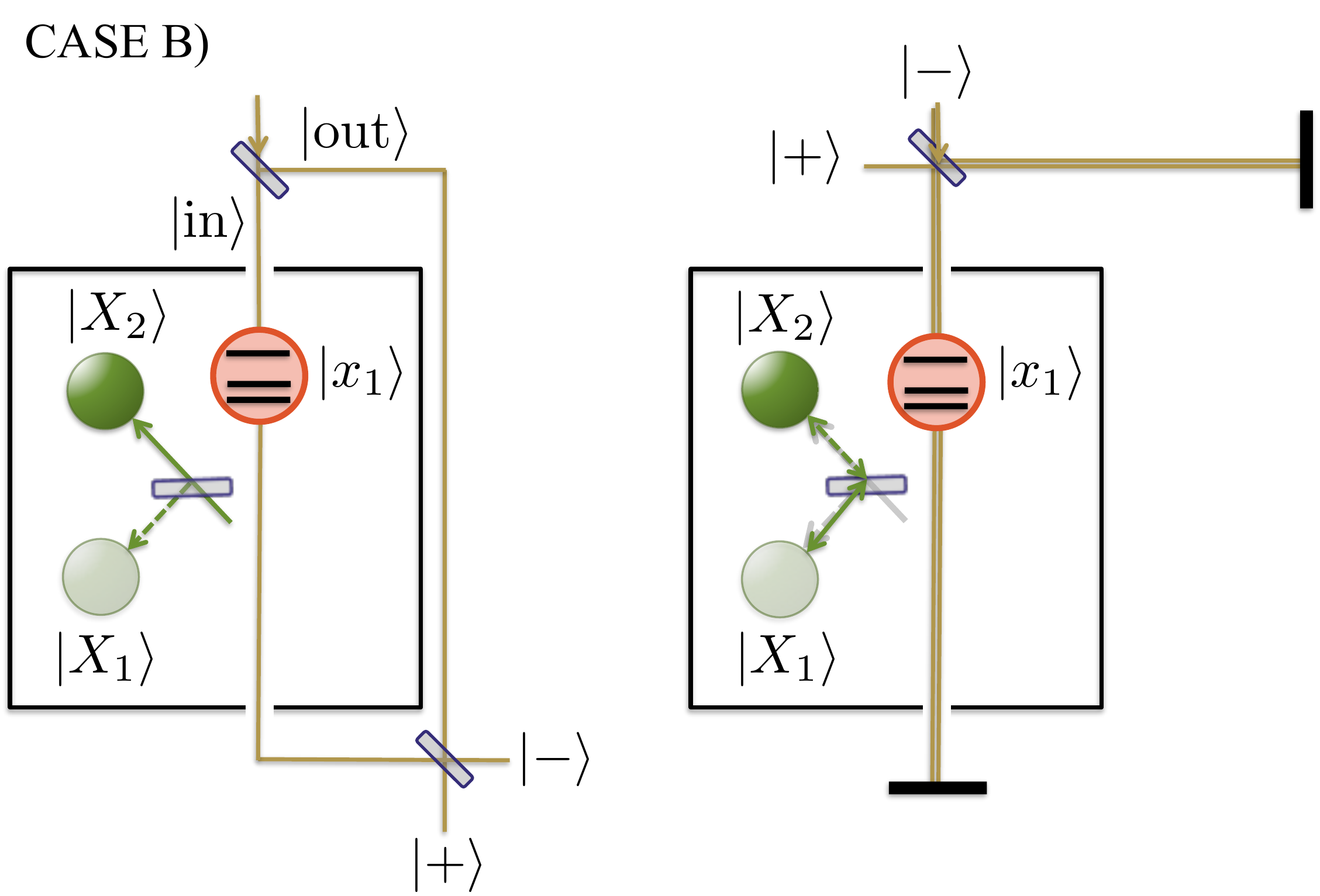}
\caption{Mass $M$ in a superposition of two distances from an atom. As in case A, an outside agent can measure the photon. The scheme on the left panel prepares the mass in one of its two orthogonal states. The scheme on the right panel measures the relative phase between the amplitudes of the mass. Both schemes yield the same outcomes as in case A. Thus, the agent cannot tell which system, the atom or the mass, is prepared in superposition and is responsible for the observed interference.\hspace*{\fill}}
\label{caseB}
\end{figure}

Case B, Figure \ref{caseB}, is fully analogous to A, but now the atom remains in a fixed state $\ket{x_1}$, while the mass $M$  is prepared in superposition of two states $\ket{X_1}$ and $\ket{X_2}$ semi-classically localised at the corresponding distances $d_1$ and $d_2$ from the atom. A photon passes through the box and the atom again acquires a phase that depends on the distances between the atom and the mass. The state of all systems after the photon leaves the box is thus analogous to eq.~\eqref{caseAfin}, where the states of the mass and the atom are simply exchanged $\ket{x_i}\leftrightarrow\ket{X_i}$. 
Overlapping the photon amplitudes on the beam splitter, Fig.~\ref{caseB} left panel, yields again the two possible outcomes with equal probabilities.  
Note that in this case the photon becomes entangled with the mass but is uncorrelated with the atom. 

As in case $A$, the outside agent can also reflect the photon amplitudes back through the box,  Fig.~\ref{caseB} right panel. 
Here, the mass $M$ goes again through the beam splitter and after the photon traverses the box it is measured in the basis $\ket{\pm}$. The two outcomes are obtained with the same probabilities $P_\pm$ as in case $A$. The outside agent can again certify coherence of the superposition state of the mass, and confirm that the interaction between the mass and the photon is unitary. 

The key observation is that the measurements on the photon do not allow the outside agent to distinguish {whether it was the atom or the mass $M$ prepared to be in superposition} --  whether the photon went through the box of case A or case B. The outside agent cannot tell whether the photon was entangled with the mass $M$ or with the atom (left panels of the figures), nor whether they measured interference of the mass or of the atom (right panels of the figures). {The individual positions of the two system are usually considered to be independent degrees of freedom (DOFs). In the standard quantum there is only one relevant DOF in the above scenarios -- the distance between the atom and the mass. The photon thus becomes entangled with the {relative distance} between the two systems (left panels) and the outside agent  interferes two superposed relative distances (right panels).} 

\paragraph{Equivalence of scenarios A and B.--}\label{general}
The examples above suggest that in the standard quantum theory the notion of superposition is not absolute. We formalise this idea and prove that scenarios in which different systems are initially prepared in a superposition state are merely different descriptions of the same process. {A crucial aspect of the argument will be that the description of measurements must be correspondingly changed as well, in order that one obtains a consistent description of the same process.}
We first prove the result for spatial superpositions and then generalise for arbitrary states. We begin with formalising the two cases to be considered. 

In case A system 1 is at an initial time $t_i$ in state $\ket{\psi_1}$, 
while system 2 is in a superposition of two states that differ by a spatial translation by $\vec {a}$ 
\be
\ket{A_i,  t_i}=N_A\ket{\psi_1}\big(\ket{\phi_2}+U_2(\vec a)\ket{\phi_2}\big),
\ee{stateA}
 where $N_A$ denotes a normalisation factor and $U_2(\vec a)$ is a unitary operator implementing spatial translations on the states of system 2, {such that if $\phi_2(\vec x_2)=\braket{\vec x_2}{\phi_2}$ then } $\bra{\vec x_2}U_2(\vec a)\ket{\phi_2}=\phi_2(\vec x_2 - \vec a)$. 
In case B, it is system 1 that is initially in a superposition 
\be
\ket{B_i, t_i}=N_B\big(\ket{\psi_1}+U_1(-\vec a)\ket{\psi_1}\big)\ket{\phi_2}, 
\ee{stateB}
with $N_B$ a normalisation factor and $U_1(-\vec a)$  a unitary operator inducing a spatial translation by $-\vec{a}$ on system 1. 
Note that the spatial distances between the two systems are the same in both cases, e.g.~if  $\ket{\psi_1}\ket{\phi_2}$ describes system 1 and 2 semi-classically localised around $\vec x_1$ and $\vec x_2$, respectively, then 
the systems are in a superposition of being separated by $\Delta \vec x=\vec x_1-\vec x_2$ and by $\Delta \vec x+\vec a$.

The probability amplitude to find the systems at time $t_f$ in some state $\ket{A_f, t_f}$ can be obtained from a propagator \cite{Ryder1996Quantum} $\mathcal{K}(\{\vec x_f\}, t_f;\{\vec x_i\}, t_i)$, where $\{\vec x\}\equiv (\vec x_1, \vec x_2)$ denotes the set of coordinates of both systems. 
In case A we are interested in the amplitude  $\mathcal{A}:=\braket{x_f, t_f}{A_i,t_i}=\int d\{\vec x\}\mathcal{K}(\{\vec x_f\},t_f;\{\vec x\},t_i)A(\{\vec x\}, t_i)$, which reads
$\mathcal{A}=N_A\int d\{\vec x\}\,\mathcal{K}\big(\{\vec x_f\},t_f; \{\vec x\}, t_i\big)\psi_1(\vec x_1)\big[\phi_2(\vec x_2)+\phi_2(\vec x_2-\vec a)\big]$

%
In  a translationally invariant theory (i.e.~with a suitably invariant action) the propagator is translationally invariant, 
which here implies
 \be
\mathcal{K}(\{\vec x_f\}, t_f; \{\vec x+\vec a\}, t_i)=\mathcal{K}(\{\vec x_f-\vec a\}, t_f; \{\vec x\}, t_i),
\ee{invariance_prop}
see also Supplemental Material. 
Linearity of $\mathcal{A}$ in the initial state and eq.~\eqref{invariance_prop} yield
\be
\begin{split}
\mathcal{A}=N_A\int d\{\vec x\}\big[\mathcal{K}&(\{\vec x_f\}, t_f; \{\vec x\}, t_i)\psi_1(\vec x_1)\\
+\mathcal{K}(\{\vec x_f-\vec a\}, t_f; \{\vec x\}, t_i)&\psi_1(\vec x_1+\vec a)\big]\phi_2(\vec x_2). 
\end{split}
\ee{equivalence_2} 

We now show that amplitude \eqref{equivalence_2} is the same as the amplitude in scenario B. {Since relative distances -- including those between the measured systems and the detectors -- cannot change under a mere change of the description, the detection amplitude now reads} 
\be
\begin{split}
\mathcal{B}=N_B&\!\int\!d\{\vec x\}\big[\mathcal{K}(\{\vec x_f\}, t_f; \{\vec x\}, t_i)\psi_1(\vec x_1)\\
+&\mathcal{K}(\{\vec x_f-\vec a\}, t_f; \{\vec x\}, t_i)\psi_1(\vec x_1+\vec a)\big]\phi_2(\vec x_2).
\end{split}
\ee{equivalence_3}
We still need to show equality of the factors $N_A, N_B$ (up to a phase). $N_{A(B)}$ can be defined by requiring that for $t_f\rightarrow t_i$ the total detection probability is 1. Since $\lim_{t_f\to t_i}{K}(\{\vec x_f\}, t_i; \{\vec x\}, t_i)=\delta(\{\vec x_f-\vec x\})$ where $\delta(\cdot)$ is the Dirac delta distribution we obtain
 $|N_A^{-2}|=|N_B^{-2}|=\int d\{\vec x\} \left| \psi_1(\vec x_1)\phi_2(\vec x_2) + \psi_1(\vec x_1)\phi_2(\vec x_2-\vec a)\right|^2$.
This concludes the proof that  scenarios A and  B are equivalent according to standard quantum formalism in any translationally invariant theory.

\paragraph{Generalisations.--}\label{general}
The above analysis directly generalises to arbitrary symmetries of dynamics. 
For a symmetry $\Lambda$ with a unitary representation 
$U(\Lambda)\ket{\psi}=\ket{\Lambda \psi}$ 
eq.~\eqref{invariance_prop} generalises to $\bra{\alpha_1}\bra{\beta_2}\mathcal{K}\ket{\Lambda\psi_1}\ket{\Lambda\phi_2} = \bra{\Lambda^{-1}\alpha_1}\bra{\Lambda^{-1}\beta_2}\mathcal{K}\ket{\psi_1}\ket{\phi_2}$ and we obtain
\be
\begin{split}
\bra{\alpha_1}\bra{\beta_2}\mathcal{K}\ket{\psi_1}\big[\ket{\phi_2}&+\ket{\Lambda\phi_2}\big]=
\big[ \bra{\alpha_1}\bra{\beta_2}\mathcal{K}\ket{\psi_1}\\
+\bra{\Lambda^{\!-1}\alpha_1}\bra{\Lambda^{\!-1}\beta_2}& \mathcal{K}\ket{\Lambda^{\!-1}\psi_1}\big]\ket{\phi_2},
\end{split}
\ee{gen_unit}
which generalises eq.~\eqref{equivalence_2} beyond translations. 
In Supplemental Material we discuss further generalisations, beyond symmetries of dynamics. 

Finally, we note that the relative nature of superpositions directly implies relative nature of entanglement.  Eq.~\eqref{gen_unit} yields
$\bra{\alpha_1}\bra{\beta_2}\mathcal{K}\big[\ket{\psi_1}\ket{\phi_2}+\ket{\Lambda\psi_1}\ket{\Lambda\phi_2}\big]=
\big[ \bra{\alpha_1}\bra{\beta_2}\mathcal{K}
+\bra{\Lambda^{\!-1}\alpha_1}\bra{\Lambda^{\!-1}\beta_2} \mathcal{K}\big]\ket{\psi_1}\ket{\phi_2}$,
which shows that measuring an entangled state in a product basis and measuring a product state in entangled basis are two different descriptions of the same scenario, differing by the choice of coordinates used to label the states of a composite system. 
 
\paragraph{Applications.--}
{One exemplary application of our result is in the context of gravity. Let us take system 1 to be a large massive body and system 2 a test-mass, e.g.~an atom, as in our thought experiment.}
When $\ket{\alpha_1}=\ket{\psi_1}$ and $\ket{\beta_2}=\ket{s(\Lambda)_2}:=\ket{\phi_2}+\ket{\Lambda\phi_2}$ eq.~\eqref{gen_unit} describes interference between amplitudes of a test-mass  in a fixed background produced by the large mass. 
Assume in addition 
 that (i) the amplitudes are orthogonal $\braket{\phi_2}{\Lambda\phi_2}\approx 0$ and  (ii) evolve only by a phase  $\bra{\psi_1}\bra{\Lambda\phi_2}\mathcal{K}\ket{\psi_1}\ket{\phi_2}\approx0$, 
 satisfied e.g.~when the evolution time is sufficiently short. 
Eq.~\eqref{gen_unit} then reads
\be
\begin{split}
\bra{\psi_1}\bra{s(\Lambda)_2}\mathcal{K}\ket{\psi_1}\ket{s(\Lambda)_2}=\\
\bra{\psi_1}\bra{\phi_2}\mathcal{K}\ket{\psi_1}\ket{\phi_2}+\bra{\Lambda^{-1}\psi_1}&\bra{\phi_2}\mathcal{K}\ket{\Lambda^{-1}\psi_1}\ket{\phi_2}.
\end{split}
\ee{example_1}
Consider now interference between states of the large mass $\ket{s(\Lambda^{-1})_1}=\ket{\psi_1}+\ket{\Lambda^{-1}\psi_1}$ when the test-mass is in a fixed state, i.e.~$\bra{s(\Lambda^{-1})_1}\bra{\phi_2}\mathcal{K}\ket{s(\Lambda^{-1})_1}\ket{\phi_2}$. Under assumptions (i), (ii)
\be
\begin{split}
\bra{s(\Lambda^{-1})_1}\bra{\phi_2}\mathcal{K}\ket{s(\Lambda^{-1})_1}\ket{\phi_2}=\\
\bra{\psi_1}\bra{\phi_2}\mathcal{K}\ket{\psi_1}\ket{\phi_2}+\bra{\Lambda^{-1}\psi_1}\bra{\phi_2}&\mathcal{K}\ket{\Lambda^{-1}\psi_1}\ket{\phi_2},
\end{split}
\ee{example_2}
which is the same as the amplitude \eqref{example_1}.
Our approach shows that two apparently different scenarios should give the same results if the underlying dynamics is translationally invariant and when (i), (ii) hold. In the first one,  
 a test-mass is prepared and measured in a superposition while a large mass is prepared and measured in a fixed state  \eqref{example_2}. In the second scenario, the test-mass is in a fixed state while the large mass undergoes interference \eqref{example_1}, ({as in cases A and B in our closed-box thought experiment).} 

There is no ambiguity in describing interference experiments with test-particles in curved spacetime~\cite{BIR82}, and thus how to compute the left hand side  (lhs) of eq.~\eqref{example_1}. However, it is generally considered that when a massive body is prepared in a superposition state of distinct locations, there is no unambiguous way to describe the lhs of eq.~\eqref{example_2}, {as this would entail the superposition of spacetime metrics, and thus pertain to the domain of quantum gravity.} Our result shows that at least for interference of macroscopically distinguishable {and static} amplitudes, no ambiguity arises: Invariance of physical laws under symmetry transformations {unequivocally} determines the amplitudes \eqref{example_1} and \eqref{example_2}  to be exactly the same.  
   
 \paragraph{Discussion}
Our result has implications for several alternative approaches aiming to describe the interface of quantum theory and general relativity, such as semi-classical gravity \cite{kiefer2012quantum} or gravitational decoherence models \cite{ref:Bassi2013}.
In semi-classical gravity the gravitational effects of macroscopic bodies on test particles are described by a classical gravitational field equal to the mean field produced by the macroscopic body.  Semi-classical gravity thus predicts amplitudes \eqref{example_1} and \eqref{example_2}  to be in general different.  If the the systems evolve by a phase we can write $\mathcal{K}\ket{\psi_1}\ket{\phi_2}=e^{i\theta(\psi_1, \phi_2)}\ket{\psi_1}\ket{\phi_2}$. Both, in semi-classical gravity and in standard quantum theory the amplitude in eq.~\eqref{example_1} (lhs) reads $e^{i\theta(\psi_1, \phi_2)}+e^{i\theta(\psi_1, \Lambda\phi_2)}$. The resulting probability is $\propto1+\cos\left(\Delta\theta\right)$ where $\Delta\theta:=\theta(\psi_1, \phi_2)-\theta(\psi_1, \Lambda\phi_2)$,  showing interference between the amplitudes, verified in numerous experiments with the earth as the large mass and neutrons \cite{cow, rauch2015book} or atoms~\cite{chu, rosi2014precision, PhysRevLett.114.013001, asenbaum2016phase, TinoFermiSchool} as the interfering test-masses. However, amplitude \eqref{example_2} (lhs) in semi-classical gravity reads $\bra{\phi_2}\mathcal{K}_{\mathrm{eff}}\ket{\phi_2}$ with  the effective propagator $\mathcal{K}_{\mathrm{eff}}$ containing gravitational field given by a mean position of  the massive body (system 1), see e.g.~\cite{Anastopoulos:2015zta}.  This amplitude is  $\propto e^{i\theta_{eff}(\psi_2)}$ and yields unit detection probability independently of the size of the superposition. As per our result the these two amplitudes must, however, be equal unless translational invariance is broken.

{The same} conclusion holds for models postulating decoherence of spatial superpositions of massive systems  where the decoherence rate depends on the superposition size and the mass of the superposed system~\cite{ref:Diosi1989, ref:Penrose1996, Anastopoulos:2015zta, Stamp2015, kiefer2012quantum, ref:Bassi2013}. 
For a sufficiently large mass of system 1, such models predict that the state $\ket{s(\Lambda^{-1})_1}$ quickly decays into mixture $\ket{\psi_1}\bra{\psi_1}+\ket{\Lambda^{-1}\psi_1}\bra{\Lambda^{-1}\psi_1}$, which in eq.~\eqref{example_2} yields a mixture of probabilities 
instead of interference. But if the mass of system 2 is sufficiently small, probabilities arising from eq.~\eqref{example_1}  remain the same as in the standard theory. Such models therefore predict the amplitudes \eqref{example_1} and \eqref{example_2} to be in general different, in violation of translational invariance. {In the context of our  closed-box thought experiments, these models allow the outside agent to distinguish cases A and B: Fast decoherence of the large mass entails $P_\pm\approx1/2$ in case B (in the Michelson-Morley configuration), while case A would remain unchanged. Therefore, such models can also be seen as necessitating the notion of an absolute position.}

{In order to satisfy translational invariance, a decoherence model cannot directly depend on the superposition size.  The only example of such a model we know of is the theory of} Kafri, Taylor and Milburn \cite{Kafri:2014zsa, Kafri:2015iha}, where decoherence depends on the superposition size of the \textit{relative distance} between a pair of systems \cite{Altamirano2016knc, Altamirano:2018KTM}. A direct extension of this model to composite systems has been shown to contradict experiments \cite{Altamirano:2018KTM}, which opens the question: {to what extent can translational invariance and classicality of gravity be simultaneously satisfied?}

Our work is in several ways complementary to the QRFs approach. It has been shown that QRFs can be fully described relative to each other, without any use of a classical reference \cite{Giacomini:2017qcovariance}. This allowed finding transformations between states, dynamics and measurements which preserve the probabilities under quantum transformations between QRFs. A general form of a Hamiltonian which remains symmetric under specific quantum transformations (superpositions of Galilean translations and superpositions of Galilean boosts) has also been found. In contrast, in our approach probabilities are conserved without constraining the allowed transformations, and invariance under ``superpositions'' of arbitrary symmetry transformations directly follows from the invariance of the dynamics under individual symmetries. However, our approach does not immediately provide the mathematical structure of the resulting transformations.

\paragraph{Conclusion.--}
Linearity of quantum theory entails that 
 measurement probabilities are invariant not only under classical transformations of coordinates, but also under more general transformations --  comprising ``superpositions'' of transformations generated by the symmetries. 
The resulting approach can be applied to processes involving macroscopic superposition states of massive bodies. It also reveals that alternative gravity theories -- aiming to preserve the features of classical general relativity at a cost of breaking the superposition principle --  can be seen as breaking translational invariance,  which is inarguably at the core of  general relativity. Finally, the approach may bring new insights into the foundations of quantum theory, as it unveils a novel limitation to the ontic interpretation \cite{Liefer:2014Quanta22} of quantum states. 
 

\paragraph{Acknowledgment} 
M.Z and F.C.~acknowledge support through Australian Research Council (ARC) DECRA grants DE180101443 and DE170100712, and ARC Centre EQuS CE170100009. T.C.R.~acknowledges support from the ARC Centre of Excellence for Quantum Computation and Communication Technology (project no.~CE170100012). This publication was made possible through the support of a grant from the John Templeton Foundation. The opinions expressed in this publication are those of the authors and do not necessarily reflect the views of the John Templeton Foundation. The authors acknowledge the traditional owners of the land on which the University of Queensland is situated, the Turrbal and Jagera people.

\bibliographystyle{linksen}
\bibliography{bibliorelsuper2_copy}

\providecommand{\href}[2]{#2}\begingroup\raggedright\begin{thebibliography}{10}

\bibitem{WeinbergGR}
S.~Weinberg, {\em Gravitation and cosmology: Principle and applications of
  general theory of relativity}.
\newblock John Wiley and Sons, Inc., New York, 1972.

\bibitem{Gabbay2006philosophy}
D.~Gabbay, P.~Thagard, J.~Woods, J.~Butterfield, and J.~Earman, {\em Philosophy
  of Physics}.
\newblock Handbook of the Philosophy of Science. Elsevier Science, 2006.

\bibitem{ref:Diosi1989}
L.~Diosi, ``Models for universal reduction of macroscopic quantum
  fluctuations,'' \href{http://dx.doi.org/10.1103/PhysRevA.40.1165}{{\em
  Phys.~Rev.~A} {\bfseries 40}, 1165 (1989)}.

\bibitem{ref:Penrose1996}
R.~Penrose, ``On gravity's role in quantum state reduction,''
  \href{http://dx.doi.org/10.1007/BF02105068}{{\em General Relativity and
  Gravitation} {\bfseries 28}, 581--600 (1996)}.

\bibitem{Anastopoulos:2015zta}
C.~Anastopoulos and B.-L. Hu, ``{Probing a Gravitational Cat State},''
  \href{http://dx.doi.org/10.1088/0264-9381/32/16/165022}{{\em Class. Quant.
  Grav.} {\bfseries 32}, 165022 (2015)}.

\bibitem{Stamp2015}
P.~C.~E. Stamp, ``Rationale for a correlated worldline theory of quantum
  gravity,'' \href{http://dx.doi.org/10.1088/1367-2630/17/6/065017}{{\em New J.
  Phys.} {\bfseries 17}, 065017 (2015)}.

\bibitem{kiefer2012quantum}
C.~Kiefer, {\em Quantum Gravity: Third Edition}.
\newblock International Series of Monographs on Physics. OUP Oxford, 2012.

\bibitem{ref:Bassi2013}
A.~Bassi, K.~Lochan, S.~Satin, T.~P. Singh, and H.~Ulbricht, ``Models of
  wave-function collapse, underlying theories, and experimental tests,''
  \href{http://dx.doi.org/10.1103/RevModPhys.85.471}{{\em Rev. Mod. Phys.}
  {\bfseries 85}, 471 (2013)}.

\bibitem{carney2018massive}
D.~Carney, P.~C. Stamp, and J.~M. Taylor, ``Massive non-classical states and
  the observation of quantum gravity: a user's manual,'' {\em arXiv preprint
  arXiv:1807.11494} (2018).

\bibitem{PhysRevD.30.368}
Y.~Aharonov and T.~Kaufherr, ``Quantum frames of reference,''
  \href{http://dx.doi.org/10.1103/PhysRevD.30.368}{{\em Phys. Rev. D}
  {\bfseries 30}, 368--385 (1984)}.

\bibitem{Rovelli:1991QRS}
C.~Rovelli, ``Quantum reference systems,''
  \href{http://dx.doi.org/10.1088/0264-9381/8/2/012}{{\em Classical and Quantum
  Gravity} {\bfseries 8}, 317 (1991)}.

\bibitem{RevModPhys.79.555}
S.~D. Bartlett, T.~Rudolph, and R.~W. Spekkens, ``Reference frames,
  superselection rules, and quantum information,''
  \href{http://dx.doi.org/10.1103/RevModPhys.79.555}{{\em Rev. Mod. Phys.}
  {\bfseries 79}, 555--609 (2007)}.

\bibitem{AngeloPhysics2011}
R.~M. Angelo, N.~Brunner, S.~Popescu, A.~J. Short, and P.~Skrzypczyk, ``Physics
  within a quantum reference frame,''
  \href{http://dx.doi.org/10.1088/1751-8113/44/14/145304}{{\em J.~Phys.~A:
  Math.~Theor.} {\bfseries 44}, 145304 (2011)}.

\bibitem{Angelo:2015PRA}
S.~T. Pereira and R.~M. Angelo, ``Galilei covariance and Einstein's equivalence
  principle in quantum reference frames,''
  \href{http://dx.doi.org/10.1103/PhysRevA.91.022107}{{\em Phys. Rev. A}
  {\bfseries 91}, 022107 (2015)}.

\bibitem{Loveridge2018}
L.~Loveridge, T.~Miyadera, and P.~Busch, ``Symmetry, Reference Frames, and
  Relational Quantities in Quantum Mechanics,''
  \href{http://dx.doi.org/10.1007/s10701-018-0138-3}{{\em Foundations of
  Physics} {\bfseries 48}, 135--198 (2018)}.

\bibitem{Giacomini:2017qcovariance}
F.~Giacomini, E.~Castro-Ruiz, and {\v{C}}.~Brukner, ``Quantum mechanics and the
  covariance of physical laws in quantum reference frames,'' {\em arXiv
  preprint arXiv:1712.07207} (2017).

\bibitem{PoundRebka:1960}
R.~Pound and G.~Rebka, ``Apparent Weight of Photons,''
  \href{http://dx.doi.org/10.1103/PhysRevLett.4.337}{{\em Phys.~Rev.~Lett.}
  {\bfseries 4}, 337--341 (1960)}.

\bibitem{cow}
R.~Colella, A.~Overhauser, and S.~Werner, ``Observation of Gravitationally
  Induced Quantum Interference,''
  \href{http://dx.doi.org/10.1103/PhysRevLett.34.1472}{{\em Physical Review
  Letters} {\bfseries 34}, 1472--1474 (1975)}.

\bibitem{TinoFermiSchool}
G.~M. Tino, \href{http://dx.doi.org/10.3254/978-1-61499-448-0-457}{``Testing
  gravity with atom interferometry,''} in {\em Proceedings of the International
  School of Physics - Enrico Fermi}, G.~M. Tino and M.~A. Kasevich, eds.,
  pp.~457--493.
\newblock 2014.

\bibitem{asenbaum2016phase}
P.~Asenbaum, C.~Overstreet, T.~Kovachy, D.~D. Brown, J.~M. Hogan, and M.~A.
  Kasevich, ``Phase Shift in an Atom Interferometer due to Spacetime Curvature
  across its Wave Function,''
  \href{http://dx.doi.org/10.1103/PhysRevLett.118.183602}{{\em Phys. Rev.
  Lett.} {\bfseries 118}, 183602 (2017)}.

\bibitem{Zych:2012}
M.~Zych, F.~Costa, I.~Pikovski, T.~C. Ralph, and C.~Brukner, ``General
  relativistic effects in quantum interference of photons,''
  \href{http://dx.doi.org/10.1088/0264-9381/29/22/224010}{{\em Classical and
  Quantum Gravity} {\bfseries 29}, 224010 (2012)}.

\bibitem{ref:Shapiro1964}
I.~I. Shapiro, ``Fourth Test of General Relativity,''
  \href{http://dx.doi.org/10.1103/PhysRevLett.13.789}{{\em Physical Review
  Letters} {\bfseries 13}, 789--791 (1964)}.

\bibitem{Ryder1996Quantum}
L.~H. Ryder, {\em Quantum Field Theory}.
\newblock Cambridge University Press, 1996.

\bibitem{BIR82}
N.~D. Birrell and P.~C.~W. Davies, {\em Quantum fields in curved space}.
\newblock No.~7. Cambridge university press, 1984.

\bibitem{rauch2015book}
H.~Rauch and S.~A. Werner, {\em Neutron interferometry: lessons in experimental
  quantum mechanics}.
\newblock Oxford University Press, USA, 2015.

\bibitem{chu}
A.~Peters, K.~Y. Chung, and S.~Chu, ``Measurement of gravitational acceleration
  by dropping atoms,'' \href{http://dx.doi.org/10.1038/23655}{{\em Nature}
  {\bfseries 400}, 849--852 (1999)}.

\bibitem{rosi2014precision}
G.~Rosi, F.~Sorrentino, L.~Cacciapuoti, M.~Prevedelli, and G.~Tino, ``Precision
  measurement of the Newtonian gravitational constant using cold atoms,''
  \href{http://dx.doi.org/10.1038/nature13433}{{\em Nature} {\bfseries 510},
  518--521 (2014)}.

\bibitem{PhysRevLett.114.013001}
G.~Rosi, L.~Cacciapuoti, F.~Sorrentino, M.~Menchetti, M.~Prevedelli, and G.~M.
  Tino, ``Measurement of the Gravity-Field Curvature by Atom Interferometry,''
  \href{http://dx.doi.org/10.1103/PhysRevLett.114.013001}{{\em Phys. Rev.
  Lett.} {\bfseries 114}, 013001 (2015)}.

\bibitem{Kafri:2014zsa}
D.~Kafri, J.~M. Taylor, and G.~J. Milburn, ``{A classical channel model for
  gravitational decoherence},''
  \href{http://dx.doi.org/10.1088/1367-2630/16/6/065020}{{\em New J. Phys.}
  {\bfseries 16}, 065020 (2014)}.

\bibitem{Kafri:2015iha}
D.~Kafri, G.~J. Milburn, and J.~M. Taylor, ``{Bounds on quantum communication
  via Newtonian gravity},''
\href{http://dx.doi.org/10.1088/1367-2630/17/1/015006}{{\em New J. Phys.}
  {\bfseries 17}, 015006 (2015)}.

\bibitem{Altamirano2016knc}
N.~Altamirano, P.~Corona-Ugalde, R.~B. Mann, and M.~Zych, ``Unitarity,
  feedback, interactionsÑdynamics emergent from repeated measurements,''
  \href{http://dx.doi.org/10.1088/1367-2630/aa551b}{{\em New J.~Phys.}
  {\bfseries 19}, 013035 (2017)}.

\bibitem{Altamirano:2018KTM}
N.~Altamirano, P.~Corona-Ugalde, R.~B. Mann, and M.~Zych, ``Gravity is not a
  pairwise local classical channel,''
  \href{http://dx.doi.org/10.1088/1361-6382/aac72f}{{\em Classical and Quantum
  Gravity} {\bfseries 35}, 145005 (2018)}.

\bibitem{Liefer:2014Quanta22}
M.~Leifer, ``Is the Quantum State Real? An Extended Review of $\psi$-ontology
  Theorems,'' \href{http://dx.doi.org/10.12743/quanta.v3i1.22}{{\em Quanta}
  {\bfseries 3}, 67--155 (2014)}.

\end{thebibliography}\endgroup

\section{Supplemental Material}

\subsection{Closed box: general case of $\Delta\varphi\neq\pi$}
\paragraph{Mach-Zehnder configuration.--} In Case A, for an arbitrary $\Delta\varphi$, when the photon leaves the box the state of all systems reads
\be
 \begin{split}
\frac{1}{2}\ket{X_1}\big[\left(\ket{x_1}+e^{i\theta}\ket{x_2}\right)e^{i\phi_c}\kout\\
+\left(\ket{x_1}+e^{i\theta+i\Delta\varphi}\ket{x_2}\right)\kin\big],
 \end{split}
\ee{A_phot_out} 
and after the photon passes through the beam splitter, as in the left panel of fig.~\ref{caseA}, we find (up to the factor $1/2\sqrt{2}$) 
\be
 \begin{split}
\ket{X_1}\big[&\big((1+e^{i\phi_c})\ket{x_1}+e^{i\theta}(e^{i\Delta\varphi}+e^{i\phi_c})\ket{x_2}\big)\ket{+}\\
+&\big((1-e^{i\phi_c})\ket{x_1}+e^{i\theta}(e^{i\Delta\varphi}-e^{i\phi_c})\ket{x_2}\big)\ket{-}\big],
 \end{split}
\ee{caseAfin_gen}
which generalises eq.~\eqref{caseAfin}. 
The detection probabilities read  $P_\pm =\big[1\pm\cos(\phi_c+\Delta\varphi/2)\cos(\Delta\varphi/2)\big]/2$ and we see that in general they depend on both $\Delta\varphi$ and $\phi_c$. The states of the atom prepared when the photon is detected in mode $\ket\pm$ read $\ket{f_\pm}\propto\big[\ket{x_1}\left(1\pm e^{i\phi_c}\right)+e^{i\theta}\ket{x_2}\left(e^{i\Delta\varphi}\pm e^{i\phi_c}\right)\big]$ and are of course in general not orthogonal. This, however, does not affect our main conclusion that the outside agent can prepare the state of the system inside the box without knowing which system has been prepared: the atom or the mass $M$. 

This is best visible when we write the state of the systems in Case B just after the photon passes through the box:  
\be
 \begin{split}
\frac{1}{2}\ket{x_1}\big[\left(\ket{X_1}+e^{i\theta'}\ket{X_2}\right)e^{i\phi_c}\kout\\
+\left(\ket{X_1}+e^{i\theta'+i\Delta\varphi}\ket{X_2}\right)\kin\big],
 \end{split}
\ee{B_phot_out}
where $\theta'= M\Delta\Phi_a t/\hbar$ is the relative phase acquired by the superposed amplitudes of the mass $M$ due to the gravitational potential of the atom $\Delta\Phi_a=Gm_a(1/d_2-1/d_1)$, therefore $\theta=\theta'$. Eq.~\eqref{B_phot_out} is  fully analogous to eq.~\eqref{A_phot_out}, with the states of the mass and the atom simply interchanged. Note that the phase acquired by the photon is explicitly the same as in Case A since it depends on the gravitational potential difference at the location of the atom between the amplitudes for which mass is at a distance $d_1$ and $d_2$ -- which are explicitly the same as in case A.

\paragraph{Michelson-Morley configuration.--} The outside agent can also reflect the photon back through the box, and after it interacts with the systems inside, measure it in a superposition basis. The sequence of events is:  the superposed state inside the box is overlapped again on the beam-splitter, which creates a new superposition between the involved states, in case A -- of the atom $\ket{x_i}$ (in case B -- of the mass $\ket{X_i}$). The photon is then reflected from the outside mirrors. The state of all system in case A now reads, up to the normalisation factor ${1}/{2\sqrt{2}}$
\be
 \begin{split}
\ket{X_1}\big[\big\{\ket{x_1}+\ket{x_2}+e^{i\theta}(\ket{x_1}-\ket{x_2})\big\}e^{i\phi_c}\kout\\
+\big\{\ket{x_1}+\ket{x_2}+e^{i\theta+i\Delta\varphi}(\ket{x_1}-\ket{x_2})\big\}\big]\kin.
 \end{split}
 \ee{A_reflect}
The photon goes through the box and all systems acquire again the  relative phases as during the first passage. This results in  in the following state when the photon leaves the box (again up to normalisation factor) 
\be
 \begin{split}
\ket{X_1}&\big[\big\{\ket{x_1}+e^{i\theta}\ket{x_2}+e^{i\theta}(\ket{x_1}-e^{i\theta}\ket{x_2})\big\}e^{i2\phi_c}\kout\\
+\big\{\ket{x_1}+&e^{i\theta+i\Delta\varphi}\ket{x_2}+e^{i\theta+i\Delta\varphi}(\ket{x_1}-e^{i\theta+i\Delta\varphi}\ket{x_2})\big\}\big]\kin.
 \end{split}
 \ee{A_reflect_pass}
Note that in this process in case A the atom and in case B -- mass $M$ -- end up going through a Mach-Zehnder interferometer. {The photon amplitudes are finally overlapped on the beam splitter outside the box, with the resulting output modes $\ket\pm\propto\kin\pm\kout$. This completes a Michelson-Morley interferometric sequence for the photon and creates the final state, which up to a factor ${e^{i\theta/2}}/{2}$ reads
\be
\ket{X_1}\sum_\pm\left(J_{1}^{\pm}(\Delta\varphi)\ket{x_1}-i e^{i\theta}J_{2}^{\pm}(\Delta\varphi)\ket{x_2}\right)\ket{\pm},
\ee{caseAfinMM_full}
where 
\be
\begin{split}
J_1^\pm(\Delta\varphi)\!=e^{i\frac{\Delta\varphi}{2}}\!\cos(\frac{\theta+\Delta\varphi}{2})\pm e^{i2\phi_c}\!\cos(\frac{\theta}{2}),\\
J_2^\pm(\Delta\varphi)\!=e^{i3\frac{\Delta\varphi}{2}}\!\sin(\frac{\theta+\Delta\varphi}{2})\pm e^{i2\phi_c}\!\sin(\frac{\theta}{2}).
\end{split}
\ee{JJ}}
The photon detection probabilities in the $\ket\pm$ basis read
\be
 \begin{split}
P_\pm(\Delta\varphi)=\frac{1}{2}\pm\frac{1}{4}\big\{ \cos(2\phi_c+\Delta\varphi)(1+\cos(\Delta\varphi))\\
+2\sin(2\phi_c+\Delta\varphi)\sin(\frac{\Delta\varphi}{2})\cos(\theta+\frac{\Delta\varphi}{2}) \big\}.
 \end{split}
\ee{general_probab_MM}
For $\Delta\varphi=\pi$ eqs \eqref{JJ} and \eqref{general_probab_MM} reduce to the expressions given in the main text,  in particular the probabilities become $P_\pm =(1\mp\sin(\theta)\sin(2\phi_c))/2$.

In a general case, the visibility of interference, when varying the phase $\phi_c$, obtained from eq.~\eqref{general_probab_MM} reads $\sqrt{\cos^4(\delta\varphi)+\sin^2(\delta\varphi)\cos^2(\theta+\delta\varphi)}/2$, where $\delta\varphi:=\Delta\varphi/2$. 
By varying the duration of the experiment (by placing the outside mirrors further/closer away from the box) the agent can vary $\theta$ and thus modulate the above visibility and even measure $\Delta\varphi$.

\subsection{Translational invariance of the propagator}\label{appendix}
 The propagator can be generally written as
\be
\begin{split}
\mathcal{K}(\{\vec x_f\}, t_f; \{\vec x_i\}, t_i)= &\\
\int_{\{\vec x_i\}}^{\{\vec x_f\}}\mathcal{D}(x(t))\,&
  e^{\frac{i}{\hbar}\int_{t_i}^{t_f}dt\,L(\{\vec x(t)\}, \{\dot{\vec x}(t)\},t)},
 \end{split}
\ee{propagator}
where $\dot a:=\frac{da}{dt}$ and $L(\{\vec x(t)\}, \{\dot{\vec x}(t)\},t$ is the total Lagrangian of the two systems.
 The integration over  paths is formally defined as $\mathcal{D}(x(t))={\prod}_{t=t_i}^{t=t_f}d\{\vec x(t)\}$ and for arbitrary but fixed vector $ \vec a$ the propagator  reads
  \be
  \begin{split}
\mathcal{K}(\{\vec x_f- \vec a\}, t_f; \{\vec x_i- \vec a\}, t_i) = &\\
 \int_{\{\vec x_i-\vec a\}}^{\{\vec x_f-\vec a\}}\prod_{t=t_i}^{t=t_f}d\{\vec x(t)\}&
  e^{\frac{i}{\hbar}\int_{t_i}^{t_f}dt\,L(\{\vec x(t)\}, \{\dot{\vec x}(t)\},t)},
 \end{split}
\ee{propagator_transl}
with the condition that $\{\vec x(t_i)\}=\{\vec x_i -\vec a\}$ and $\{\vec x(t_f)\}=\{\vec x_f -\vec a\}$. Shifting the integration variable to $\vec x'(t)=\vec x(t)+\vec a$ for all $t\in[t_i, t_f]$, and for both systems, the rhs of eq.~\eqref{propagator_transl} reads
\be
 \int_{\{\vec x_i\}}^{\{\vec x_f\}}\prod_{t=t_i}^{t=t_f}d\{x(t)\}e^{\frac{i}{\hbar}\int_{t_i}^{t_f}dt\,L(\{\vec x(t)-\vec a\}, \{\dot{\vec x}(t)\},t)}.
\ee{integral_shift}
A theory is translationally invariant if $L(\{\vec x(t)-\vec a\}, \{\dot{\vec x}(t)\},t)=L(\{\vec x(t)\}, \{\dot{\vec x}(t)\},t)$ in which case
 expression~\eqref{integral_shift} is explicitly equivalent to $\mathcal{K}(\{\vec x_f\}, t_f; \{\vec x_i\}, t_i) $  and thus
\be
\mathcal{K}(\{\vec x_f- \vec a\}, t_f; \{\vec x_i- \vec a\}, t_i)=\mathcal{K}(\{\vec x_f\}, t_f; \{\vec x_i\}, t_i). 
\ee{transl-inv_prop}

\subsection{Beyond symmetries of dynamics}
Any two normalised states can be transformed into each other via some unitary $V$. If $V$ is not a symmetry of dynamics, in general we have $\mathcal{K}^V:=V^\dagger\mathcal{K}V\neq\mathcal{K}$. 
However, the probability amplitudes are still invariant, in the sense that
\be\bra{b}\mathcal{K}\ket{a}+\bra{b}\mathcal{K}\ket{Va}=\bra{b}\mathcal{K}\ket{a}+\bra{V^\dagger b}\mathcal{K}^V\ket{a}\ee{superpos_actions}
 where $\ket{Va}:=V\ket a$. Eq.~\eqref{superpos_actions} means that superposition of arbitrary states can also be seen as relative -- the difference is that for some scenarios the system is then described as evolving under different actions, here $\mathcal{K}^V$ and $\mathcal{K}$, ``in superposition''. An example would be a system prepared in a superposition of states at different values of some external potential. When the initial state is expressed by a single amplitude, the system is described as be evolving in superposition under actions with different values of the potential.

\end{document}